\documentclass{ws-procs975x65}

\def\beq{\begin{equation}}
\def\eeq{\end{equation}}

\begin{document}

\title{Bayesian hierarchical modelling of weak lensing --- the golden goal}
\author{Alan Heavens$^*$, Justin Alsing, Andrew Jaffe, Till Hoffmann}

\address{ICIC (Imperial Centre for Inference and Cosmology)\\
Imperial College, South Kensington\\
London SW7 2AZ, U.K.\\
$^*$E-mail: a.heavens@imperial.ac.uk}

\author{Alina Kiessling} 

\address{Jet Propulsion Laboratory, California Institute of Technology, \\4800 Oak Grove Drive, \\
Pasadena, CA 91109, USA\\}

\author{Benjamin Wandelt}

\address{Institut d'Astrophysique de Paris, UMR CNRS 7095, \\Universit\'{e} Pierre et Marie Curie, 98bis boulevard Arago\\
75014 Paris, France}

\begin{abstract}
To accomplish correct Bayesian inference from weak lensing shear data requires a complete statistical description of the data. The natural framework to do this is a Bayesian Hierarchical Model, which divides the chain of reasoning into component steps. Starting with a catalogue of shear estimates in tomographic bins, we build a model that allows us to sample simultaneously from the the underlying tomographic shear fields and the relevant power spectra ($E$-mode, $B$-mode, and $EB$, for auto- and cross-power spectra). The procedure deals easily with masked data and intrinsic alignments. Using Gibbs sampling and messenger fields, we show with simulated data that the large (over 67000-)dimensional parameter space can be efficiently sampled and the full joint posterior probability density function for the parameters can feasibly be obtained. The method correctly recovers the underlying shear fields and all of the power spectra, including at levels well below the shot noise.\end{abstract}

\keywords{Cosmology; statistics; gravitational lensing}

\bodymatter


\section{The risks of future experiments}

Cosmology has often been described in recent years as moving into the {\em precision era}, where error bars at percent level are commonplace in the cosmic microwave background. In future, this precision will increase enormously in related fields, such as large-scale structure, weak gravitational lensing, and, in due course, 21cm studies, with ambitious observational programmes connected with Euclid, the Large Synoptic Survey Telescope, and the Square Kilometre Array.  The surveys associated with these facilities will deliver exquisite precision, which will help to answer many of the major cosmological questions of the decade, such as the nature of Dark Energy, possible beyond-Einstein gravity, macroscopic extra dimensions, and the masses and hierarchy of neutrinos.  However, with this precision comes risk: the small statistical error bars that come from the large datasets involved will lead to potentially far-reaching and erroneous conclusions if the {\em accuracy} of the results is not correspondingly high.  To achieve this requires very good control of systematic effects at all stages of the experiment, including in the statistical analysis of the data.   In this report, we focus on weak gravitational lensing, and propose a rigorous statistical framework that can encapsulate essentially everything the data tell us, allowing for accurate and suitably precise inferences to be made, with full propagation of errors.

\section{The goal of scientific inference}

Let us focus on the scientific method. A very general description of an experiment is that there is some prior information $I$, some new data $d$, and one or more models $M$ with parameters $\theta$.  W assume we wish to infer the parameters of a given model in the light of the new data collected by the experiment.  From a Bayesian perspective, everything we know is encapsulated in the posterior probability,
\[
p(\theta | d, I) \propto p(d | \theta, I ) p(\theta | I)
\]
where we have used Bayes' theorem to write it in terms of the likelihood and the prior, and the dependence on the model has been suppressed.  This is it; one can if desired form {\em estimates} of the parameters from this, but that is not necessarily useful, and one can from this form Bayesian credible intervals for the parameters.  The posterior is the desired outcome of the experiment --- it is in a sense the `golden goal' of statistical analysis.

\subsection{Computing the posterior}

Inevitably, for anything other than a very simple experiment, the posterior probability is not an analytic function, and cannot be computed directly, and the most common general technique is to draw {\em samples} from the posterior (or sometimes the likelihood).  Asymptotically, methods such as Markov Chain Monte Carlo will sample the target distribution with a density of samples that is proportional to the target, so with a sufficient number of samples, the chain characterises the target to whatever accuracy is required, although there may be  computational constraints.

So, is there any reason not to try to construct the posterior? The answer is basically no --- if you can do it, you should.  However, for some situations this may be computationally a very demanding task.  However, recent theoretical advances and the increase in computational power mean that many problems for which previously this golden goal was unachievable are now soluble.  The posterior itself is complicated, but it may be broken down into a hierarchy of elements, each of which we do understand and can sample from.  This forms what is called a `Bayesian Hierarchical Model' (BHM). The state-of-the-art has reached a stage where one of the most demanding cosmological probes is now accessible to a rigorous treatment --- weak gravitational lensing on a cosmic scale.

\section{Weak gravitational lensing}

Weak gravitational lensing\cite{Munshi2008} distorts the shapes and sizes of galaxy images, and the measured shape of an individual image can be used as a very noisy estimate of the spin-weight 2 shear field induced by lensing.
A Bayesian Hierarchical Model for lensing could in principle be built from the raw image pixel data all the way to inferences about cosmology,  but here we restrict ourselves to a subset of the analysis process, taking as a starting point estimates of the shear at the locations of individual galaxies.  This does not propagate all errors fully, but is a major first step forward.  We also assume that we have photometric redshifts for the galaxies, and bin them in both photo-z and pixels on the sky, in sufficient numbers that we can assume the errors (which are dominated by intrinsic ellipticity dispersion) are gaussian.  Thus we have pixelised noisy shear maps in some number of tomographic bins, represented by the data, written in terms of the true shear maps $s$ and the noise $n$ (characterised by a covariance matrix $N$) as $d=s+n$ ($d$ represents the complete data vector, including pixels in all tomographic bins).  The statistical properties of the maps are set by various power spectra (auto- [within a bin], and cross-, in both $E$- and $B$-mode, and $EB$ cross-powers).  These power spectra are controlled in a physical model by cosmological parameters, and we have a choice of performing a cosmology-independent characterisation of the data, or inferring cosmological parameters. Thus we have as unknown parameters either: 
\begin{itemlist}
\item True\ shear\ maps $s$, plus power\ spectra\ $C^{XY}_{\alpha\beta}; \ X,Y=E {\rm\ or\ } B; \ \alpha,\beta = {\rm bin} $,
\item or true\ shear\ maps $s$, plus cosmological parameters $\theta$.
\end{itemlist}
There are advantages to doing both, but let us concentrate initially on the first. The power spectra all enter into $C=\langle s\,s^T\rangle$, we may parametrise $C$ by band powers. We have a relatively large number of parameters about $C$ to infer, and a very large number ($\sim 10^5-10^7$) of true pixelised shear values.  Thus it is a very high-dimensional problem.  Statistically we are interested most in $p(C|d)$, so would marginalise over $s$, but if one is interested in map-making, then one can obtain $p(s|d)$ by marginalising over $C$.

\subsection{Weak lensing Bayesian Hierarchical Model}

\begin{figure}[ht]
\begin{center}
\includegraphics[width=1.in]{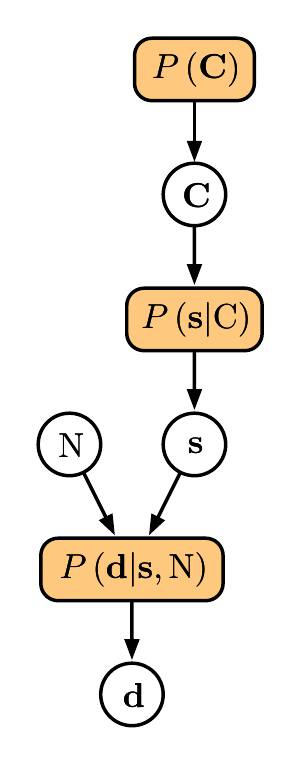}
\end{center}
\caption{Simple Bayesian Hierarchical Model for weak lensing. $C$ represents the lensing power spectra, $s$ the true shear fields, $d$ the data, and $N$ the noise covariance.}
\label{BHM}
\end{figure}

A weak lensing BHM is shown in Fig.~\ref{BHM}; Alsing et al.\cite{Alsing2016} give further details.  The generative model for the data consists of drawing samples of $C$, from which shear maps can be drawn, since the conditional probability $p(s|C)$ is known, and noise be added since the conditional probability $p(d|s,N)$ is also known.  BHMs lend themselves naturally to Gibbs samplers, which sample alternately from the power spectra and the map:
\begin{align}
&C^{i+1} \leftarrow P(C|s^i) \nonumber \\
&s^{i+1} \leftarrow P(s|C^i, d, N).
\end{align}
Interestingly, for gaussian fields, the distributions (conditioned on the observed data $d$) can be written down: $p(C|s)$ is an inverse-Wishart distribution; $p(s|C, d, N)$ is a gaussian with a mean given by the Wiener-filtered map, 
$
d_{\rm WF} = (C^{-1}+N^{-1})^{-1}N^{-1}d
$
with covariance $(C^{-1}+N^{-1})^{-1}$.  However, there is a difficulty in that these matrices are huge (up to say $10^7 \times 10^7$ elements).  Unless they are diagonal, they cannot be handled at all.  Now, $C$ is diagonal in the harmonic domain, and $N$ is diagonal in the pixel domain, but there is no basis in which both of them are diagonal. This problem has been elegantly solved\cite{Elsner2013} by the introduction of a {\em messenger field}.   

\subsection{Results from simulations}

We applied the algorithm\cite{Alsing2016} sketched above to simulated noisy, masked 2-bin tomographic shear maps generated using SUNGLASS simulations\cite{Kiessling2011}, inferring  $1940$ power spectrum parameters. The shear maps contain $128\times 128$ pixels, bringing the total number of parameters in the inference task to $67476$.

\begin{figure}[ht]
\begin{center}
\includegraphics[width = 12cm]{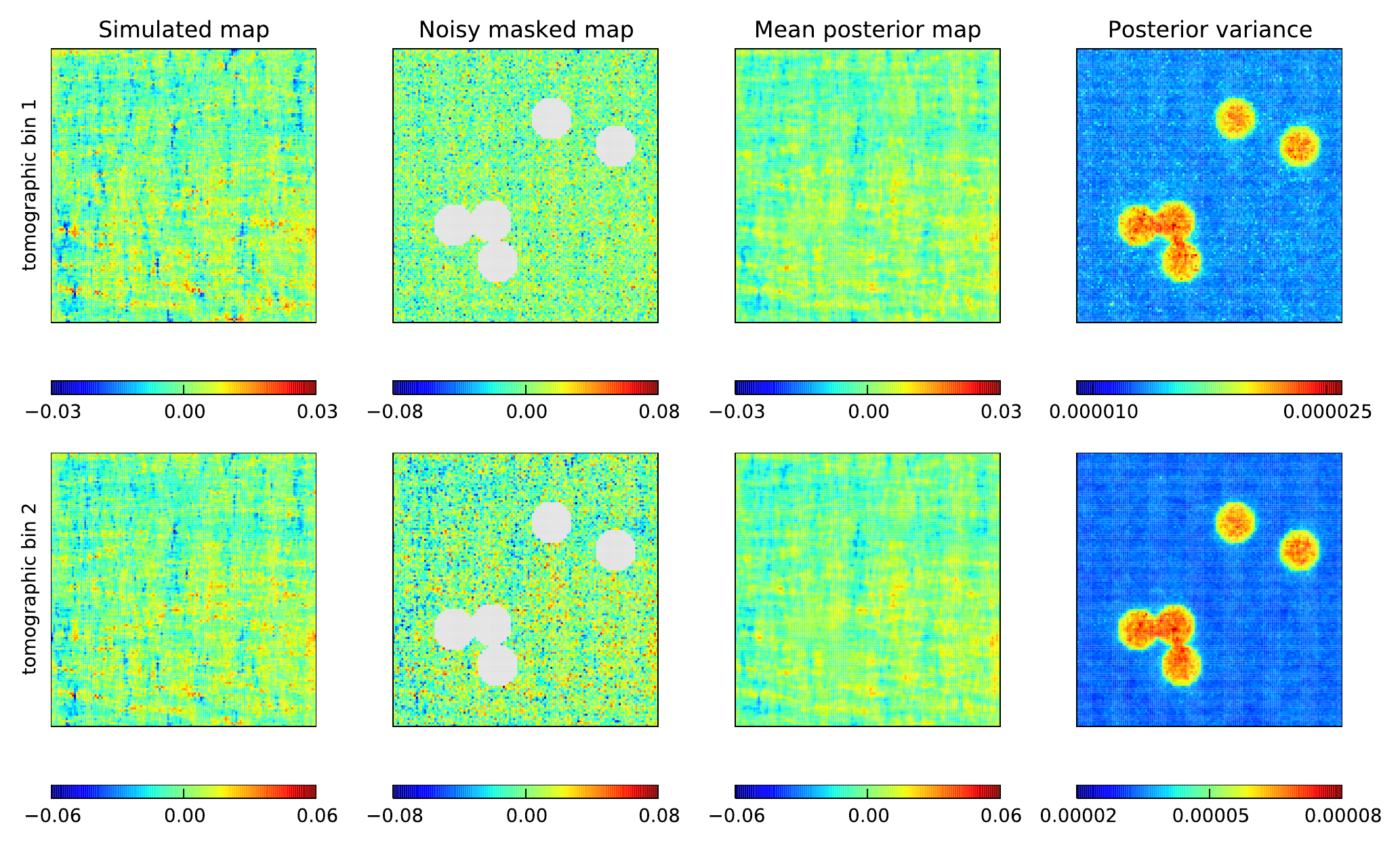}
\end{center}
\caption{Tomographic maps of the $\gamma_1$ component of the shear for the noiseless simulated shear maps (far left), noisy masked simulated maps (second from left), mean posterior maps (third from left) and the posterior variance (far right). The mean posterior maps recover most of the structure from the simulated shear maps. Note that inference is made about the field in masked regions, but the posterior variance in those regions is significantly higher than in unmasked regions.}
\label{fig:maps}
\end{figure}

\begin{figure}[ht]
\begin{center}
\includegraphics[width = 12cm]{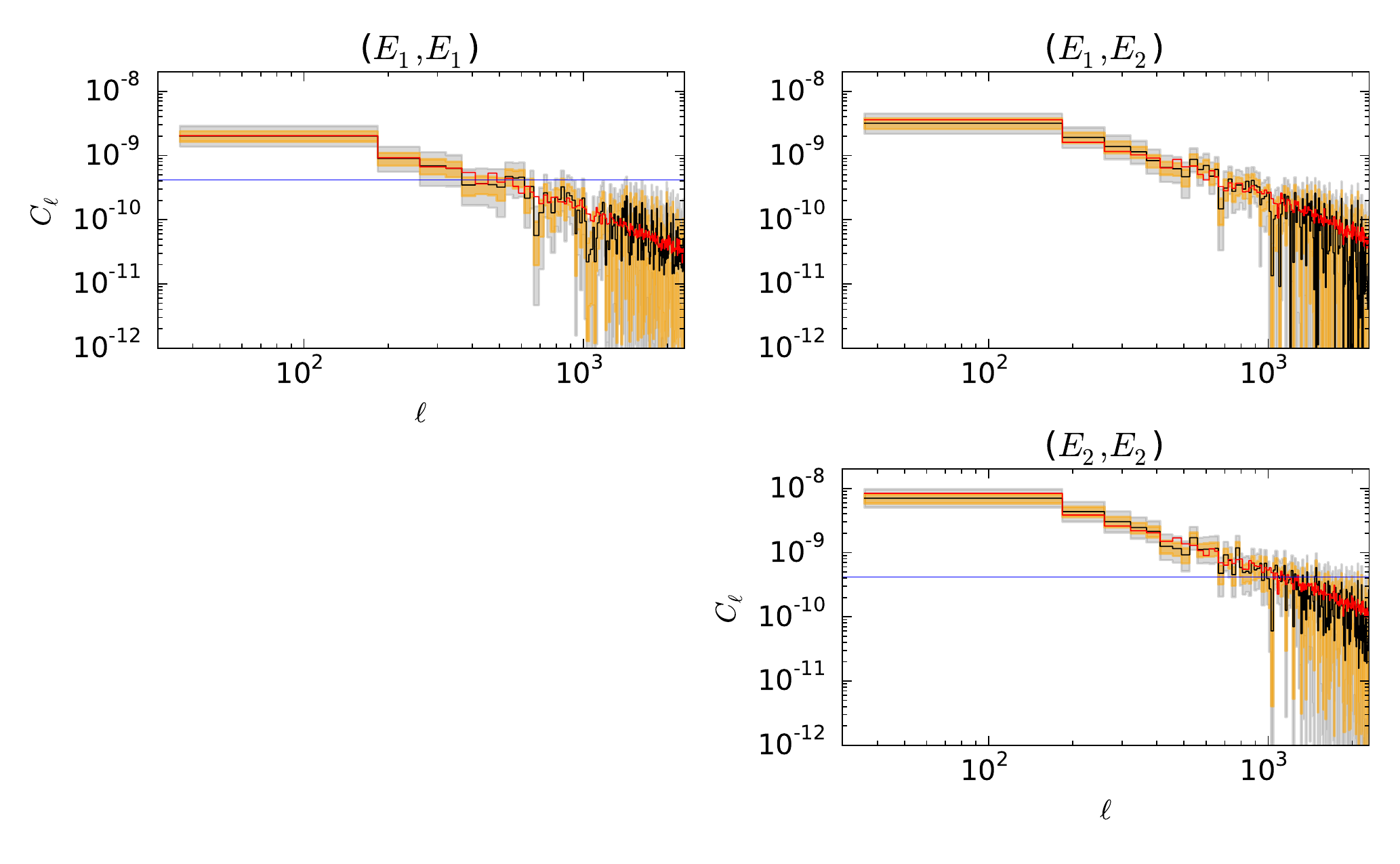}
\end{center}
\caption{Recovered $E$-mode tomographic shear power spectra: the orange (grey) bands indicate the $68\%$ ($95\%$) credible regions, the black lines show the posterior means, the red lines show the estimated band powers from the noiseless, mask-less simulated shear catalogue and the horizontal blue lines show indicate the mean ellipticity-noise level.}
\label{fig:ee}
\end{figure}

We ran three Gibbs chains with independent starting points, obtaining $3.6M$ samples, each taking $0.5$s on a desktop. Convergence was determined with a Gelman-Rubin statistic $r < 1.1$ for the marginal distributions of all parameters. 
Fig. \ref{fig:maps} shows the results in the map domain.
Fig. \ref{fig:ee} shows the recovered $E$-mode tomographic (auto and cross) power spectra for the two tomographic bins.

\section{Cosmological parameter inference and intrinsic alignments} 

An alternative to using the power spectra as parameters is to use cosmological parameters, with a Boltzmann code to generate the  lensing power spectra.  This has some advantages, in that the number of parameters is reduced, and some sampling issues in the very low S/N regime may be avoided.  An additional advantage is that it may be possible to include non-gaussianity in the conditional distributions.  The disadvantage is that the analysis becomes model-dependent, and the power spectrum approach gives a neat encapsulation of the statistical properties of the data at the 2-point level, regardless of the origin of the signal.  The other advantage of using the power spectrum approach is that it is in principle possible to do the cosmological parameter inference as a second step, using the power-spectrum chains.  This way, different models for intrinsic alignments of galaxies (a potential source of significant systematic error in weak lensing) may be investigated without rerunning chains.  They can be included in a post-processing step as an addition to the theoretical power spectrum (rather than as a source of correlated noise, which would be much harder to deal with).  Using power spectra, one could in principle include higher-order, non-gaussian terms (bispectrum, trispectrum), but in the general case this is unfeasible, as there is an extremely large number of parameters unless they are regularised by imposing a physical model.  In this case the correct procedure will be to use a gaussian likelihood, as it is the maximum entropy distribution given a mean and covariance, and is thus in some sense the least informative and most conservative assumption.  Both approaches should be undertaken.

\section{Conclusion}

We have shown here how a Bayesian Hierarchical Model can be used to generate samples efficiently from the posterior distribution --- the `golden goal' of statistical analysis ---  for tomographic weak lensing shear maps, and a multitude of lensing power spectra, inferring parameters in a $\sim 10^5$-dimensional space.  It solves many of the awkward problems of cosmic shear, such as how to treat the mask (pixel variances are simply set to infinity), and how to include intrinsic alignments (simply as an additional signal power contribution).

\end{document}